\documentclass{article}

\PassOptionsToPackage{numbers,compress}{natbib}
\usepackage[final,sglblindworkshop]{neurips_2026}
\workshoptitle{Machine Learning and the Physical Sciences}

\usepackage[utf8]{inputenc}
\usepackage[T1]{fontenc}
\usepackage{amsmath,amsfonts,booktabs,graphicx,microtype,url}
\usepackage[hidelinks]{hyperref}

\title{Probing Cosmology with Diffusion Models:\\
Memorization, Generalization, and Parameter Estimation}

\author{%
  Jiaming Pan\\
  Department of Physics and Leinweber Institute for Theoretical Physics\\
  University of Michigan, Ann Arbor, MI 48109\\
  \texttt{jiamingp@umich.edu}
  \And
  Nicholas Kern\\
  Department of Physics and Leinweber Institute for Theoretical Physics\\
  University of Michigan, Ann Arbor, MI 48109\\
}

\begin{document}

\maketitle

\begin{abstract}
Diffusion models are increasingly used as surrogates for expensive cosmological
simulations, but simple metrics like reconstruction fidelity do not guarantee a surrogate can be
trusted for scientific inference. To explore this, we train diffusion models on H{\sc i} fields from the
CAMELS IllustrisTNG simulations and test how three properties change with
training-set size: memorization of individual training examples, agreement with
one- and two-point statistics (the pixel-value distribution and the power
spectrum), and estimation of cosmological parameters from the generated fields.
At fixed model size, generative models trained with smaller, less diverse training data produced memorized samples.
While this is a well known phenomenon, we show that this leads to miscalibrated parameter densities for cosmological inference tasks.
At fixed size, there is a training set size with enough information content to enable a model to generalize, at which point posterior density estimation with neural surrogates can be well-calibrated.
We demonstrate that a neural surrogate can therefore pass visual, one-point and power-spectrum convergence tests while still resulting in overconfident
parameter estimation in a cosmological inference pipeline. Thus, tests of model generalization should be an additional check for inference pipelines adopting neural surrogates.
\end{abstract}

\section{Introduction}

Diffusion models have been shown to be highly adept at learning high-dimensional data distributions and avoiding the curse of dimensionality
\citep{sohldickstein2015deep,song2019generative,ho2020denoising},
and are increasingly being adopted as surrogates for expensive scientific forward models \citep[e.g.][]{Legin2024}.
Cosmological inference is one such application, requiring costly simulations of large-scale structure growth.
A learned surrogate could reduce this cost, but its
usefulness depends on what it learns from a finite simulation set.

We examine when diffusion models move from memorizing training set samples to true generalization: the ability to generate novel, unseen fields that still lie on the data generating manifold.
We then measure whether these fields are actually drawn from the
correct set of cosmological parameters: a necessary check for performing the inverse problem of backing-out cosmological parameters. We first measure training-set similarity, then
compare one-point and two-point distributions to assess sample fidelity, and finally test accurate recovery of the cosmological
parameter posteriors.
This is a subtle point.
A model trained on a small training set can give the right power spectrum while memorizing its training data, yet return poorly calibrated confidence intervals for a parameter inference task.
To test this, we train a supervised regression head that maps fields to their cosmological parameters, which sets our ground truth, and apply it to fields from conditional diffusion models trained on different training set sizes.
Similar to experiments on natural images, we find that generalization occurs rapidly at a training set size that grows with model capacity.
This suggests an underlying learning dynamics for generative models that could eventually enable a priori prediction for when a model of a given capacity will generalize on a training set with a given sample diversity, which we do not address in this work.

\paragraph{Related work.}
While the memorization-to-generalization transition is a well-understood phenomenon for natural images \citep{kadkhodaie2024generalization,gu2025memorization,zhang2024emergence}, this has yet to be studied in detail for cosmological datasets and parameter inference settings.
Related theory analyzes when reverse diffusion collapses onto individual training
examples as a function of sample size and data dimension
\citep{biroli2024dynamical}.
\citet{zhang2024emergence} also study class-conditional generation, measuring
reproducibility between models given the same noise and class label rather than
recovery of the conditioning variable. In cosmology, diffusion models have been
assessed using field-level summary statistics \citep{mudur2022can} and applied to
conditional field emulation and parameter inference
\citep{mudur2023cosmological,mudur2025diffusionhmc}. Here we compare
training-set novelty, summary-statistic agreement, and cosmological parameter
recovery across training set sizes.

\section{Setup}

\paragraph{Data.}
We use 2D H{\sc i} fields from the IllustrisTNG simulations in the CAMELS
Multifield Dataset \citep{Villaescusa-Navarro2022}.
The CAMELS cosmological parameters are
$\boldsymbol{\theta}=
(\Omega_m,\sigma_8,A_{\rm SN1},A_{\rm AGN1},A_{\rm SN2},A_{\rm AGN2})$, determining the amount of matter in the universe, its clustering, and star formation feedback on the gas distribution.
A single field in our training data is a $128\times128$ pixel slice of a $128^3$ H{\sc i} grid, from the LH and CV sets at $z=0$, 1 and 2 (unconditional models) or the LH set (conditional models).
Our
training sweep at fixed model size spans $N_{\rm 2D}=2^6$--$2^{15}$ and, for the unconditional models, uses every 8th slice of each $128^3$ snapshot cube (16 slices per cube).
For supervised settings, we reserve 32 simulations as the test set; they are excluded from training of the conditional models and of the regressor.
Maps are log-transformed and normalized with an asymmetric $\tanh$ rescaling before training. No data augmentation is applied, except where noted.

\paragraph{Models and training.}
We adopt three attention-UNets \citep{ronneberger2015unet} with increasing size, spanning channel widths $(16,32,64)$, $(32,64,128)$, and
$(64,128,256)$, denoted UNet-64, UNet-128, and UNet-256.
Continuous conditioning is applied to the diffusion models by encoding the ($\Omega_m,\ \sigma_8$) parameter vector into the embedding space through cross-attention.
With 10\% label dropout they also allow classifier-free guidance of scale $w$; we use plain conditional sampling ($w=1$) unless noted.
We train all models for an equal number of 200k mini-batch steps with the AdamW optimizer (learning rate=$10^{-4}$, weight decay=0.01, batch size=32) on an NVIDIA A40 GPU.
We use a 500-step squared-cosine noise schedule and a $v$-prediction denoising loss \citep{salimans2022progressive}.
At inference time, we use an accelerated 50-step DPM solver \citep{lu2022dpmsolver}, which shows agreement with a standard DDPM solver.

\paragraph{Criterion 1: training-set similarity.}
We represent each map in a principal component analysis (PCA) feature space, $\mathbf z$.
For a generated map $x_j$, we compute its maximum cosine similarity to the
training maps $y_i$,
\[
s_j =
\max_i
\frac{
\mathbf z(x_j)^\top \mathbf z(y_i)
}{
\|\mathbf z(x_j)\|_2\,\|\mathbf z(y_i)\|_2
},
\]
where $\mathbf z$ is the projection onto the 6,697 principal components that hold
98\% of the variance of the $2^{15}$ training set.
Following the use of feature-space similarity to study memorization \citep{zhang2024emergence},
we define the ``sample novelty'' score $G=\Pr[s_j<\tau]$.
For each training set, $\tau$ is the 95th percentile of the similarities between each training map and its nearest other training map: in other words, a generated map counts as memorized if it is closer to a training map than 95\% of training maps are to each other.
Thus, small $G$ indicates high training-set memorization whereas large $G$ indicates a lack of memorization.
Note that $G$ measures only sample novelty: random noise would also score high, so it is necessary but not sufficient for generalization. 

\paragraph{Criterion 2: distributional agreement.}
We compare one-point distributions (histograms of normalized pixel values) and
2-point distributions (the azimuthally averaged 2D power spectrum of each mean-subtracted field) between diffusion generated and real fields, plotting the generated mean against the real training-set median and central 68\% and 95\% ranges. These ranges describe variation among real fields,
not uncertainty on their mean.

\paragraph{Criterion 3: cosmological parameter density estimation.}
Ultimately we care about inferring cosmological parameters, and specifically we are concerned about accurately calibrated posterior distributions, $p(\theta|x)$.
We define a simplified test of whether our generative models could serve as surrogates in an inference pipeline.
Separate from the diffusion model, we use a pre-trained, frozen image encoder \citep[VGG16;][]{Simonyan2015VeryDE} to project the training samples into a rich embedding space. We then train a simple MLP regression head in a supervised manner to map from cosmological fields to the full parameter vector $\boldsymbol{\theta}$, achieving an $R^2=0.90$ for $\Omega_m$ (which acts as a ground truth for parameter recovery).
We then take our conditional diffusion models trained with different training sample sizes and draw samples conditioned on 32 cosmologies $(\Omega_m,\sigma_8)$ from our held-out set.
For each held out sample, we generate 64 fields with different random noise seeds.
We pass these fields through the encoder and regression head to recover their cosmology.
We measure the empirical posterior coverage as the fraction of the 32 held-out cosmologies whose requested $\Omega_m$ falls inside the central interval of the 64 estimates, as a function of the interval's nominal confidence level. These intervals are the spread of the regressor's predictions across fields.
We bootstrap resample these estimates to build an ensemble for the estimated posterior coverages. As a reference we apply the same test to real fields, using the 128 slices of each held-out simulation; their 68\% intervals contain the true value for 27 of the 32 simulations, above nominal.

\begin{figure}[t]
  \centering
  \includegraphics[width=0.66\linewidth]{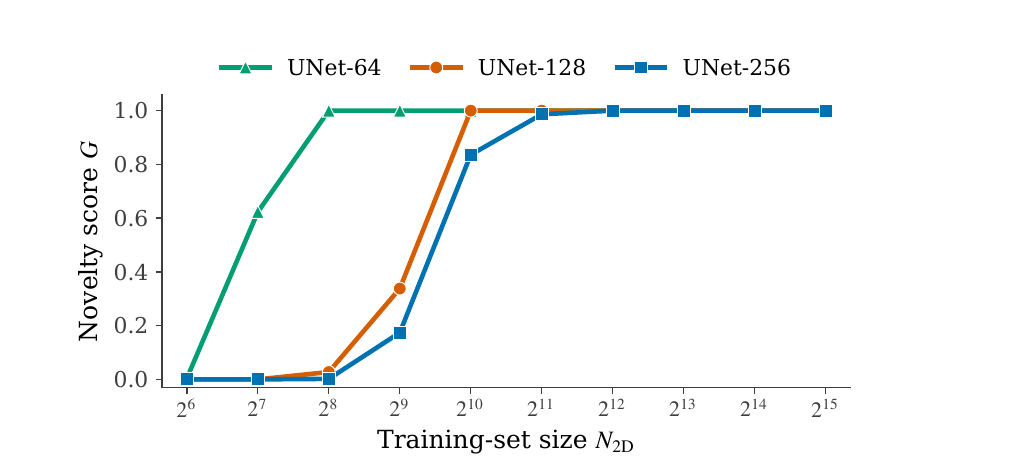}
  \caption{
  \textbf{Transition from training-set memorization to generalization.}
  Novelty score $G$ for three UNet widths. Each model reaches generalization at a different data set size, influenced by the competing pressure to overfit the data at large model size and the diversity of data in the training set.
  }
  \label{fig:transition}
\end{figure}

\section{Results}

\subsection{A data-dependent transition in training-set memorization}

The fraction of samples classified as novel increases with training-set size
(Fig.~\ref{fig:transition}). This transition occurs at larger training-set sizes
for larger UNet models, which can be understood as the inductive bias of the model to overfitting.
Because the budget is fixed, smaller training sets are seen for many more epochs, so the location of the transition partly reflects this choice.
For UNet-128, $G$ increases from 0.338 at $N_{\rm 2D}=512$ to 1 at 1024; UNet-256 increases from 0.174 to 0.836 over the same range, reflecting similar overall behavior observed for natural image datasets \citep{zhang2024emergence}.
The critical point at which a model generalizes is due to competing pressures of the intrinsic data diversity, the model's inductive bias, and the denoising diffusion loss, whose exact minimizer reproduces the training data \citep{biroli2024dynamical}.
An ultimate goal would be to develop an information theoretic description of when a model with a set of inductive biases $\mathcal{B}$ will generalize when trained on a dataset with a given data diversity $\mathcal{D}$, which would aid researchers when deploying generative models for arbitrary scientific tasks.

\subsection{Distributional agreement}

\begin{figure}[t]
  \centering
  \includegraphics[width=\linewidth]{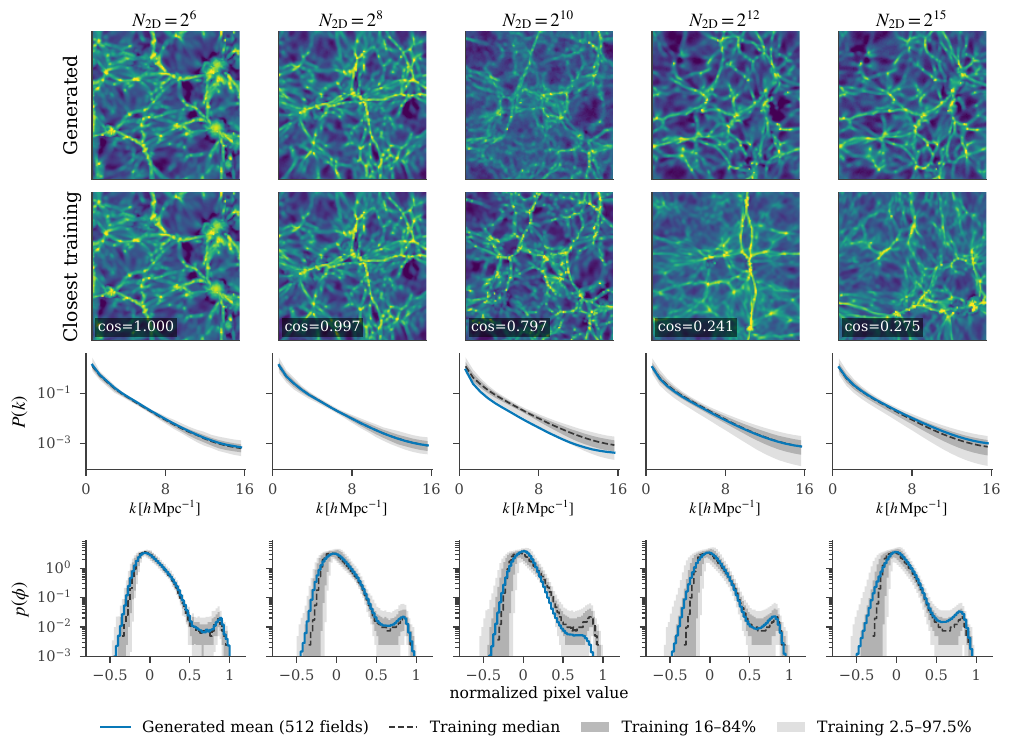}
  \caption{
  \textbf{Reconstruction fidelity and training set memorization at fixed model size.}
  Top and second rows: after training, we generate one new field and find its closest match in its training set, with their pixel cosine similarity (cos), plotted as a function of training set size. Third and bottom rows: generated mean power spectrum and
  one-point distribution (512 fields; blue), training data median (dashed), and
  central 68\% (dark grey) and 95\% (light grey) training set ranges.
  }
  \label{fig:nearest_u128}
\end{figure}

Figure~\ref{fig:nearest_u128} compares generated fields with their closest
training fields for five UNet-128 models. At $N_{\rm 2D}=2^6$ and $2^8$,
generated one-point distributions and power spectra agree well with the real
training fields, but the generated samples nearly reproduce individual training set
examples (pixel cosine similarities of 1.000 and 0.997), as do the samples as a
whole (median nearest-training similarity 0.9997 and 0.994 over 512 samples).

At $N_{\rm 2D}=2^{10}$, the generated fields are more distinct from the training
set, while the power spectrum shows a scale-dependent discrepancy and the
one-point distribution lacks the highest-density pixels.
At $2^{12}$ and $2^{15}$, this model achieves broad generalization and the power spectrum agreement improves. Distributional agreement therefore worsens around the transition and improves beyond it.


\begin{figure}[t]
  \centering
  \includegraphics[width=0.88\linewidth]{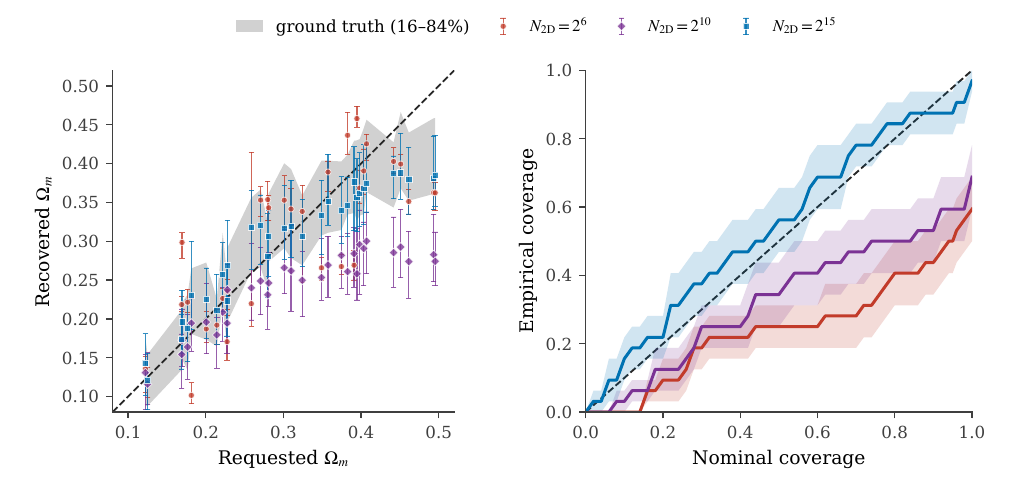}
  \caption{
  \textbf{Parameter estimation and posterior coverage calibration.}
  Left: true vs. recovered $\Omega_m$ parameter. Grey shaded shows the $1\sigma$ confidence interval ground truth of our regression head, while points show recovered parameter values and $1\sigma$ uncertainties at held-out $\Omega_m$ values in the test set.
  Right: empirical coverage versus nominal coverage, with shaded regions showing 68\% interval from bootstrap sampling. A memorized model (red) is unbiased on average but has errorbars that are too small (is overconfident). A generalized model (blue) is unbiased and its coverage approaches that of the ground truth.
  }
  \label{fig:recovery}
\end{figure}

\subsection{Parameter density estimation depends on training set size}

At $N_{\rm 2D}=2^6$, the conditional model reproduces training fields instead of
generating new ones at the held-out cosmologies (Fig.~\ref{fig:recovery}). At
$w=1$ the recovery looks unbiased: the median recovered $\Omega_m$ matches the
real-field value (median bias $-0.001$). But the 64 draws at a cosmology are
nearly the same training field, the intervals are about 2.5 times narrower than
for real fields, and the nominal 68\% and 95\% intervals contain the requested
value for only 9 and 16 of the 32 cosmologies, compared with 27 and 32 for real
fields. At $w\le0.5$ the recovered $\Omega_m$ is instead biased low, and for every
guidance scale we tested coverage stays at or below 17 of 32.

The response and coverage improve with more training data. At
$N_{\rm 2D}=2^{10}$, the generated fields are no longer copies, but they fail the
other checks at once: their pixel-value distribution deviates most from real
fields of the five sizes, they lack about a third of the large-scale power, the
recovered $\Omega_m$ is too low at large values, and the intervals contain the
requested value for 15 and 19 cosmologies, even though the separate unconditional
UNet-128 sweep reaches $G=1$ at this size. From $2^{12}$, coverage approaches that
of real fields (25 and 24 of 32 at 68\% for $2^{12}$ and $2^{15}$); at $2^{15}$
it cannot be distinguished from real fields at 68\%, but at 95\% it is still below
them.


\section{Discussion}

Training set memorization, agreement in one- and two-point statistics, and
parameter estimation test different properties of a diffusion
surrogate. Power spectra of generated samples can agree while the fields nearly reproduce training samples, so a surrogate can
look right by eye and by summary statistic while still memorizing its training data. 
The impact this has on cosmological inference is that memorized surrogates can lead to posterior density estimates that are too narrow (i.e. are overconfident).
Even with different noise seeds, a memorized diffusion model will return a narrow spread in cosmological fields that under-covers the true cosmological parameters, leading to overconfidence in the resultant posterior coverage.
Ultimately, this work aims to promote the importance of quantifying generalization for neural surrogate performance in cosmology alongside more standard metrics such as image fidelity and summary statistic reproduction.

\noindent
Future work will aim to develop techniques for data augmentation and selection to improve generalization with a fixed training set, and to develop predictive tests of generalization given pre-defined notions of training data diversity and model capacity.

\emph{Caveats--}
These conclusions rest on one simulation suite (CAMELS IllustrisTNG; the LH and CV sets for the unconditional sweep, LH only for the conditional models), one field (H{\sc i}), one cosmological parameter, and one training set realization per configuration.
We focus on $\Omega_m$ because parameter recovery is intrinsically weaker for the other parameters ($R^2=0.68$ for
$\sigma_8$ and at most 0.42 for the feedback parameters, against 0.90 for $\Omega_m$).



\begin{ack}
We thank the CAMELS collaboration for making their simulations publicly
available. This research was supported in part through computational resources
and services provided by Advanced Research Computing at the University of
Michigan, Ann Arbor. We acknowledge the following individuals for helpful discussions during our work: Qing Qu, Yueying Ni, Dragan Huterer, and the anonymous peer reviewers of the NeurIPS ML-for-Physical-Sciences workshop.
\end{ack}

\bibliographystyle{unsrtnat}
\bibliography{references}

\clearpage
\appendix
\setcounter{figure}{0}
\renewcommand{\thefigure}{A\arabic{figure}}

\end{document}